\documentstyle[psfig,aps,prl,amssymb]{revtex}
\begin{document}
\title{Volume Elements of Monotone Metrics on the $n \times n$ Density
Matrices as Densities-of-States for Thermodynamic Purposes. II}
\author{Paul B. Slater}
\address{ISBER, University of
California, Santa Barbara, CA 93106-2150\\
e-mail: slater@itp.ucsb.edu,
FAX: (805) 893-7995}

\date{\today}

\draft
\maketitle
\vskip -0.1cm

\begin{abstract}
We derive explicit expressions for the volume elements of both the minimal
and maximal monotone metrics over the $(n^{2} -1)$-dimensional convex set of
$n \times n$ density matrices for the cases $n=3$ and 4.
We make further progress for the specific $n = 3$ 
maximal-monotone case, by taking the limit of a certain ratio
of integration results, obtained using
an orthogonal set of eight coordinates. By doing so, we  
find remarkably simple {\it marginal} probability distributions
based on the corresponding volume element, which we then use for
thermodynamic purposes.
We, thus, find a spin-1 analogue of the Langevin function.
In the fully general $n=4$ situation, however, we are impeded in making
similar progress by the inability to diagonalize a $3 \times 3$
Hermitian matrix and thereby obtain an orthogonal set of coordinates to
use in the requisite integrations.
\end{abstract}


\hspace{1.4cm} Keywords: quantum statistical thermodynamics, spin-1 systems,
Langevin function, Brillouin function,

\hspace{3.2cm} monotone metrics, harmonic mean, prior probability, entangled
spin-1/2 systems

\tableofcontents

\section{INTRODUCTION}
In this communication, we report a number of results pertaining to a
certain quantum-theoretic
 model of the thermodynamic properties of a system comprised 
 of spin-1 particles, in particular, a small number of them.
Our approach can be contrasted with the
standard (Jaynesian \cite{jaynes,balian,buzek}) one, which, ``in some
respects can be viewed as semiclassical [and] can presumably be
justified when the number of the constituent particles is large --- in
which case the random phases can be averaged over \ldots However,
in the case of a small system \ldots there seems to be no
{\it a priori} reason for adopting the conventional mixed state
approach'' \cite{brody}.
Vigorous criticisms of the ``orthodox information-theoretic foundations
of quantum statistics''
have been expressed by Band and Park in an extended series of paper
 \cite{park}. Park \cite{parkalone} himself later wrote
that ``the details of quantum thermodynamics are presently
unknown'' and ``perhaps there is more to the concept of thermodynamic
equilibrium than can be captured in the canonical density operator
itself.'' Additionally, Lavenda \cite{lavenda} argued (as detailed in
sec.~\ref{comparative})
 that there are deficiencies --- from a probabilistic
 viewpoint --- with the
``Brillouin function'' (used in the study of ferromagnetism), which
is yielded by the standard methodology \cite{brody}. Contrastingly,
Lavenda asserts that the ``Langevin function'' is free from such defects.
(Aharoni \cite[pp. 83,97,98]{aharoni}, citing Yatsuya {\it et al}
 \cite{yatsuya}, in
support, reaches similar conclusions under an assumption of complete
spatial
isotropy and arbitrariness of the direction of the applied field.)
We have previously found \cite[App. I]{slater1} (cf. \cite{brody}) that
 for the spin-${1 \over 2}$ systems, the form of analysis we will pursue
here for the spin-1 systems, does, in fact, yield the Langevin function. Thus,
our methodology appears to be not subject to the criticism of Lavenda.
One of the principal results below will be a spin-1 version 
(Fig.~\ref{LLang}) of the (spin-${1 \over 2}$) Langevin
function. One possible application of these results is to quantum
chromodynamics, where one can regard the antiscreening of the Yang-Mills
vacuum as paramagnetism for the gluons, which are charged particles
of spin 1 \cite{gross,nielsen}

In the context of {\it entangled} quantum systems --- to which we seek to
apply our analytical approach
in sec.~\ref{spinspin} --- it has been argued
\cite{horo1} that ``there {\it are} situations where the Jaynes principle
fails''. ``The difficulties in understanding of the Jaynes inference scheme
are due to the fact that the latter is just a {\it principle} and it was
not derived within the quantum formalism'' \cite{horo2}.
Friedman and Shimony \cite{friedman}, in a classical rather than quantum
context,
 claim to have ``exhibited an anomaly in Jaynes'
maximum entropy prescription,'' cf. \cite{shimony1,shimony2,shimony3}.

\section{SPIN-1 SYSTEMS} \label{spin1systems}
\subsection{Background}
Bloore \cite{bloore}
 had studied the geometrical structure of the eight-dimensional convex set of
spin-1 density matrices, which he denoted,
\begin{equation} \label{blooremat}
\rho = \pmatrix{a & \bar{h} & g\cr
h & b & \bar{f}\cr
\bar{g} & f & c\cr}, \qquad a, b, c \in {\Bbb{R}}, \qquad 
 f, g, h \in {\Bbb{C}}
\end{equation}
(we incorporate the  notation of \cite{bloore} into ours).
In
the (MATHEMATICA \cite{wolfram}) computations upon which
we rely, a sequence of
 transformations --- suggested by this work of
 Bloore  --- is implemented, leading to the full separation of the
transformed
 variables \cite{miller}.
 In particular, we make use of a four-dimensional version
 $(r,\theta_{1},\theta_{2},\theta_{3})$ of spheroidal
coordinates \cite{dassio}.

 In recent years,  there have been
several studies \cite{hub1,hub2,braun,twamley,petz0}
 concerning the assignment of certain
 natural
Riemannian metrics to sets of density matrices, such as  (\ref{blooremat}).
 These various 
metrics can all
be considered to be particular forms of {\it monotone}
 metrics \cite{petzsudar,hasegawa,lesniewski}.
(Contrastingly,
 there is only {\it classically},  as shown by Chentsov \cite{chentsov},
 a {\it single} monotone metric --- the one associated with the {\it Fisher
information}. Morozova and Chentsov \cite{moroz}
 sought to extend this work to the quantum domain, while
 Petz and Sudar \cite{petzsudar} further developed
 the line of analysis.)
Of particular interest are the maximal monotone metric (of the {\it left}
 logarithmic
derivative) --- for which the reciprocal (${{\mu + \nu} \over 2}$)
 of the ``Morozova-Chentsov function''
\cite{petz} is the
{\it arithmetic} mean --- and the
minimal (Bures) monotone metric (of the {\it symmetric} logarithmic derivative)
--- for which the reciprocal $( {{2 \mu \nu} \over {\mu + \nu}})$ of the
 Morozova-Chentsov function is the {\it harmonic} mean.
Also of considerable
 interest is the Kubo-Mori metric \cite{petz0} --- the reciprocal
$({{\log{\mu} -\log{\nu}} \over {\mu - \nu}})$  of the
 Morozova-Chentsov
function of which is the {\it logarthmic} mean -- but we do not study
 it here, if for no other reason than that
it appears to be computationally intractable for our purposes.
In any case, we should note that in \cite{slater3},
relying upon a variant of a statistical
 test devised by Clarke \cite{clarkejasa},
 the Kubo-Mori metric
was found to give rise to a {\it less} noninformative prior distribution
than the maximal monotone metric, which itself was shown to be {\it most} 
noninformative (but only if --- due to nonnormalizability --- the pure
and some collection of nearly pure states, were eliminated from
consideration).
We will make use of the Morozova-Chentsov functions in deriving 
 (by adopting certain
work of Dittmann \cite{dittmann}) our formulas for the volume elements
of the minimal and monotone metrics.

\subsection{Principal Results}

The first set of questions which we wish to address is whether one can
{\it normalize}
 the {\it volume} elements of these two (minimal and maximal) metrics over
the {\it eight}-dimensional convex set of spin-1 density matrices, so as to
obtain (prior) probability distributions.
In the maximal case (as also in its spin-1/2 counterpart
 \cite[eqs. (43)-(46)]{slater1}),
the answer is strictly {\it no}, since we encounter {\it divergent} integrals.
Nevertheless --- similarly to the analysis in the spin-1/2 case --- we are
able to normalize the volume element of the maximal monotone
metric over a {\it subset} of
the entire convex set, omitting the pure and some collection
of nearly pure states. Then, by taking the limit (in which the subset
approaches the complete set) of a certain ratio, 
 obtain a
{\it lower}-dimensional {\it marginal}
probability distribution.
In this manner, we have been able to assign the probability distribution
(cf. (\ref{end})),
\begin{equation} \label{principal}
{15 (1-a) \sqrt{a} \over 4 \pi \sqrt{b} \sqrt{c}},
\end{equation}
to the {\it two}-dimensional simplex spanned by the diagonal entries of
(\ref{blooremat}).
Of course, by the trace condition on density matrices, we have that
$a+b +c =1$. The asymmetry under exchange that is
 evident between $a$ and either
 $b$ or $c$ --- but lacking between
$b$ and $c$ themselves --- in 
(\ref{principal}), is attributable to
the specific sequence of transformations, suggested by the work of
Bloore \cite{bloore},
 employed below, following
 his notational and analytical scheme. Consequently,
it is quite natural to intrepret the variable
$a$ (despite the particular ordering of rows and columns
 used by Bloore in (\ref{blooremat}),
whose notation we have adopted from the outset of our analysis)
 with the {\it middle}
 level of the three-level system
(the one inaccessible to a spin-1 photon, due to its masslessness),
 and $b$ and $c$
with the other two (accessible to a {\it two}-level system).
  With another, but equally valid sequence of
transformations,
we could have interchanged the roles of these variables in (\ref{principal}). 
(Let us also note that without the factor, $(1-a)$, (\ref{principal})
would be proportional to a member of the Dirichlet family of
distributions \cite{ferguson}.)

\begin{figure} 
\centerline{\psfig{figure=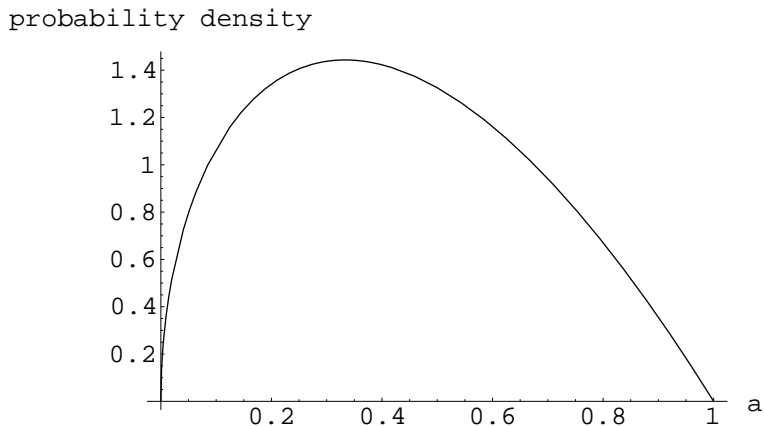}}
\caption{Univariate marginal probability distribution over the diagonal
entry $a$}
\label{Marginala}
\end{figure}

\begin{figure}
\centerline{\psfig{figure=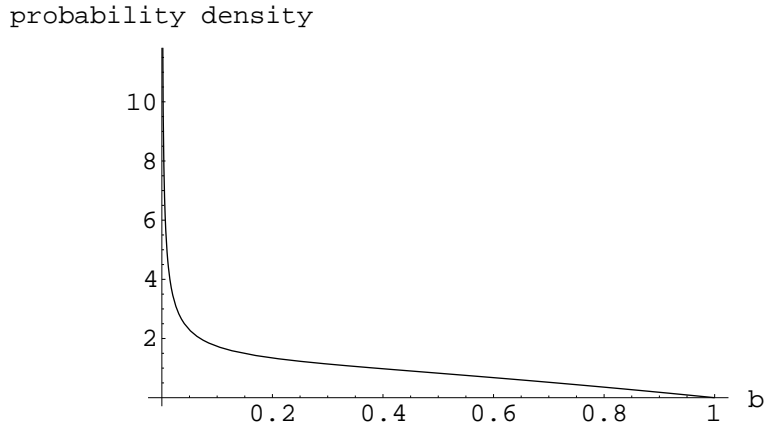}}
\caption{Univariate marginal probability distribution over the diagonal 
entry $b$}
\label{Marginalb}
\end{figure}

The {\it univariate} marginal distributions of (\ref{principal}) are
(Fig.~\ref{Marginala}) (a member of the family of beta distributions),
\begin{equation} \label{marginala}
{15 (1-a) \sqrt{a} \over 4},
\end{equation}
having
 a peak at ${a = {1 \over 3}}$, at which the probability density equals
${5 \over 2 \sqrt{3}} \approx 1.44338$
and (Fig.~\ref{Marginalb}),
\begin{equation} \label{marginalb}
 {15 (1-b) (1+3 b) \over 32 \sqrt{b}},
\end{equation}
(and similarly for the diagonal entry $c$).
The 
 expected values for these distributions
 are $\langle a \rangle = {3 \over 7}$, and
 $\langle b \rangle =
\langle c \rangle = {2 \over 7}$, so $\langle a \rangle + \langle b
\rangle + \langle c \rangle = 1$. Also, $\langle a b \rangle = {2 \over 21},
\quad \langle a^{2} \rangle = {5  \over 21},
 \quad \langle b^{2} \rangle = \langle c^{2} \rangle = {1 \over 7}$.

\subsection{Thermodynamic Analyses Based on Two {\it Diagonal} Hamiltonians}
\subsubsection{The first diagonal Case ($\lambda_{8}$)}
Let us now consider the observable, one of a standard set (but, due
to the particular sequence of transformations we employ,
suggested by the work of Bloore \cite{bloore}, we take
the liberty of harmlessly
permuting the first and third rows and columns of the usual form of
 presentation) of
 eight Hermitian generators of
$SU(3)$ \cite{arvind,byrd},
\begin{equation} \label{gellmann}
\lambda_{8}= {1 \over \sqrt{3}} \pmatrix{-2 & 0 & 0\cr
0 & 1 & 0 \cr
0 & 0 & 1 \cr},
\end{equation}
which might possibly function as the Hamiltonian of the spin-1 system.
The expected value of (\ref{gellmann})
 with respect to (\ref{blooremat}) is $ \langle \lambda_{8}
\rangle = \mbox{Tr}
 (\rho \lambda_{8}) = {{1-3 a} \over \sqrt{3}}$.
Multiplying the univariate marginal probability distribution
 (\ref{marginala}) by the {\it Boltzmann factor} 
$\exp(-\beta \langle \lambda_{8} \rangle)$ and integrating over $a$
from 0 to 1, we obtain the (``weak  equilibrium'' \cite{park})
 partition function (cf. \cite{slatergibbs}),
\begin{equation} \label{part}
 Q(\beta) =  
{1 \over {16 \beta^{5/2}}} (5 \mbox{e}^{-{{\beta \over \sqrt{3}}}}
 (6 \sqrt{\beta}
(\mbox{e}^{\sqrt{3} \beta} -3^{{1 \over 4}}
(2 \beta +\sqrt{3}) \sqrt{\pi} \mbox{erfi} (3^{{1 \over 4}} \sqrt{\beta}))),
\end{equation}
where erfi represents the imaginary error function ${\mbox{erf} (iz)
\over i}$. (The error function erf($z$) is the integral of the Gaussian
distribution, that is, $ {2 \over \sqrt{\pi}} \int_{0}^{x} \mbox{e}^{-t^{2}}
\mbox{d}t$.)
In ``strong equilibrium,'' {\it zero} prior probability is assigned to those
density matrices which do {\it not} commute with the Hamiltonian,
 while in ``weak equilibrium,'' this requirement is not imposed \cite{park}.

\begin{figure} 
\centerline{\psfig{figure=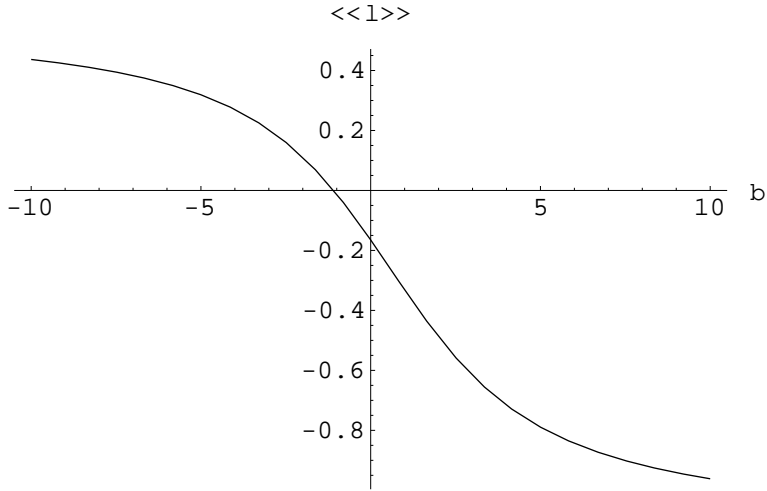}}
\caption{Expected value of $\langle \lambda_{8} \rangle$ as a function
of the inverse temperature parameter $\beta$}
\label{expvallambda}
\end{figure}

In the conventional manner of thermodynamics,
 we compute the expected value 
($\langle \langle \lambda_{8} \rangle \rangle$) of $\langle \lambda_{8}
\rangle$ as
 $- {\partial \log{Q(\beta)} \over \partial  \beta}$. The result is
plotted in Fig.~\ref{expvallambda}.

The expected value for $\beta=0$, corresponding to infinite temperature,  is
$ \langle \lambda_{8} \rangle = -{1 \over 5 \sqrt{3}} \approx -.11547$.
``Two physical conditions must be met in order for negative temperatures to
arise: the subsystem possessing the negative temperature must be well
insulated thermally from the other modes of energy storage of the complete
system, and the subsystem energy levels must be bounded from above and below.
Thus a normally populated state with probabilities proportional to
$e^{-\beta \epsilon_{i}}$ can be inverted by reversal of the order of the
energy levels while populations remain intact because there is no
convenient energy sink available. Examination of the entropy gives further
insight to the idea of negative temperatures \ldots When the energy levels 
are bounded from above as well as below, however, zero entropy can occur
for both minimum and maximum energy \ldots No particular difficulties arise
in the logical structure of statistical mechanics or of thermodynamics as
a consequence of these  negative-temperature systems''
\cite[pp. 131, 132]{robertson}.

Besides (\ref{gellmann}), the only diagonal member of
the standard set of eight Hermitian generators of $SU(3)$ is
\begin{equation} \label{gellmann2}
\lambda_{3} = J_{3}^{(1)} = \pmatrix{0 & 0 & 0 \cr
                      0 & 1 & 0 \cr
                      0 & 0 & -1 \cr},
\end{equation}
where $J_{3}^{(1)}$ is used to
 denote a particular angular momentum matrix \cite[p. 38]{bied}.
This can be viewed as the spin-1 analogue of the Pauli matrix,
\begin{equation} \label{Pauli}
\sigma_{3} = 2 J_{3}^{({1 \over 2})} =\pmatrix{1 & 0 \cr
                      0 & -1 \cr}.
\end{equation}
Use of this observable ($\sigma_{3}$)
 in conjunction with the volume element (inversely proportional to the
${3 \over 2}$-power of the determinant of the $2 \times 2$
density matrix) of the maximal
monotone metric for spin-${1 \over 2}$ systems, has led to the partition
function,
\begin{equation} \label{spin1/2part}
Q(\beta) ={\sinh{\beta} \over \beta},
\end{equation}
yielding as the expected value, the negative of the {\it Langevin} function 
\cite{tusy},
\begin{equation} \label{Langevin}
-{\partial \log{Q(\beta)} \over \partial \beta} =
{1 \over \beta} -\coth{\beta}.
\end{equation}

\subsubsection{The second diagonal case ($\lambda_{3}$)}

\begin{figure}
\centerline{\psfig{figure=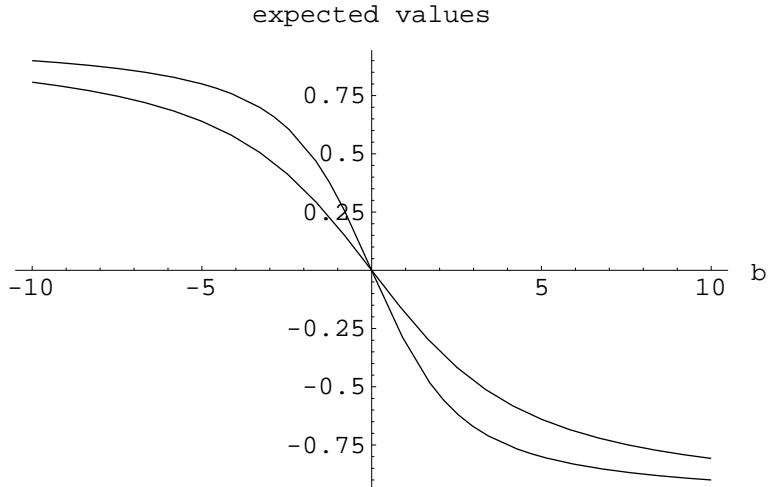}}
\caption{Expected value of $\langle \lambda_{3} \rangle$ as a function of
the inverse temperature parameter $\beta$, along with the expected value 
predicted by
the
 (steeper-at-the-origin) negative of the Langevin function (\ref{Langevin})}
\label{LLang}
\end{figure}

Now, we have additionally that
 $\langle \lambda_{3} \rangle = \mbox{Tr}(\rho \lambda_{3}) = b-c$.
Then, using a combination of symbolic and numerical integration, we have
 conducted a thermodynamic
analysis for $\lambda_{3}$,  analogous to that above 
 ((\ref{part}) and Fig.~\ref{expvallambda}) for
$\lambda_{8}$.
In Fig.~\ref{LLang}, we display the expected value ($\langle \langle
\lambda_{3} \rangle \rangle$) of $\langle \lambda_{3} \rangle$, 
obtained numerically, along
with that predicted by the negative of the Langevin function (\ref{Langevin}).
The two curves are rather similar in nature, with the negative
of the Langevin function assuming, in general, greater {\it absolute}
expected values for a given $\beta$.

\subsection{Comparative properties of the Brillouin and
 Langevin functions} \label{comparative}

\subsubsection{Critique of Lavenda}

The usual use of the Langevin function is to describe the thermodynamic
behavior of noninteracting  particles with {\it nonquantized} spin. In this
same paradigm, the {\it Brillouin} function, that is
 $\tanh{\beta}$, is employed for
spin-${1 \over 2}$ particles \cite{tusy}. 
(Its spin-1 counterpart is ${2 \sinh{\beta} \over
 1 + 2 \cosh{\beta}}$.) However, Lavenda
 \cite[pp. 193]{lavenda} has argued that, in contrast to the Langevin
function,
the Brillouin function lacks a proper probabilistic foundation, since its
generating function cannot be derived as the Laplace transform of a prior
probability density''. He also writes \cite[p. 198]{lavenda}:
``Even in this simple case of the Langevin function \ldots
we have witnessed a transition from a statistics dictated by the 
central-limit theorem, at weak-fields, to one governed by extreme-value
distributions, at strong-fields. Such richness is not possessed by the
Brillouin function, for although it is almost identical to the Langevin
function in the weak-field limit, the Brillouin function becomes independent
 of the field in the strong-field limit. In the latter limit, it would imply
complete saturation which does not lead to any probability distribution.
This is yet another inadequacy of modeling ferromagnetism by a Brillouin
function, in the mean field approximation.''
Additionally, Lavenda asserts \cite[p. 20]{lavenda} that the Langevin function
``has empirically been targeted as providing a good description of hysteresis
curves in ferromagnetic materials when the field due to interdomain coupling
is added to the actual internal field to produce an effective field.
This effective field is analogous to the Weiss mean field experienced
by individual magnetic moments within a single domain. For a sufficiently
large
interdomain coupling parameter, an elementary form of hystersis loop has
 been observed
in the modified Langevin function. Moreover, since the generating function
must be expresed as a Laplace transform of a prior distribution, in order
to make physical as well as statistical sense, this rules out certain other
candidates like the Brillouin function. For particles of spin-${1 \over 2}$,
the generating function  would be proportional to the hyperbolic cosine, and
the hyperbolic cosine cannot be expressed as a Laplace transform of a
prior distribution.''

\subsubsection{The Jiles-Atherton theory}

In his claims that the Langevin function provides a good description of
certain magnetic phenomena, Lavenda refers to the work of Jiles and 
Atherton \cite{jiles1,jiles2}.
``The Jiles-Atherton theory is based on considerations of the dependence
of energy dissipation within a magnetic material resulting from changes
in its magnetization. The algorithm based on the theory yields five
computed model parameters, $M_{s},a,\alpha,k$ and $c$, which represent
the saturation magnetization, the effective domain density, the mean
exchange coupling between the effective domains, the flexibility of domain
walls and energy-dissipative features in the microstructure, respectively
\ldots The model parameter $a$ is derived from an analogy to the Langevin
expression for the anhysteretic magnetization $M_{an}$ as a function of both
temperature $T$ and field $H$ for a paramagnet \ldots However, in the
Jiles-Atherton theory, the spin entity $\langle m \rangle$ is not an atomic
magnetic moment $m = n \mu_{b}$, where $\mu_{B}$ is the Bohr magneton,
as in the original Langevin expression. Rather, it represents the moment
from a mesoscopic collections of spins that we refer to as an
`effective domain;' each `effective domain' possesses a collective
magnetic moment $\langle m \rangle$. These effective domain entities
may or may not correspond to actual magnetic domains'' \cite{lewis}.

\subsubsection{Critique of Brody and Hughston}

 Brody and Hughston \cite{brody,brody2}  propose the
use of the negative of the
Langevin function, that is (\ref{Langevin}), for the internal energy of 
a (small) system of spin-${1 \over 2}$ particles in thermal equilibrium.
Brody (personal communication) has suggested that ``after all, standard
(Einstein's) approach to quantum statistical mechanics does seem to work
for bulk objects, so there seems to be some kind of `transition'
from micro to macro scales''. ``We note that in the case of the quantum
canonical ensemble the heat capacity for this [spin-${1 \over 2}$]
system is nonvanishing at zero temperature. Since it is known in the case
of many bulk substances that the heat capacity vanishes as zero temperature
is approached, it would be interesting to enquire if a single electron
possesses a different behaviour, as indicated by our results''
\cite{brody}.

\subsubsection{Critique of Aharoni}

In sec.~5.2, entitled ``Superparamagnetism'' (that is,
 the ``phenomenon of the loss
of ferromagnetism in small particles'') of his recent text
\cite[pp. 97, 98]{aharoni}, Aharoni writes:
``A single particle of such a small size cannot be made or handled.
Experiments are therefore carried out on an ensemble of particles,
which in most cases have a wide distribution of particle sizes.
Such particles would give rise to a superposition of Langevin functions
with different values of $\mu = M_{s} V$ in the argument, and the
measured curve could not possibly look like the Langevin function \ldots
With improved techniques for producing very small particles, their size
distribution has become narrow enough for a pure Langevin function
[citing \cite{yatsuya}] to be
observed \ldots 
The calculation [of the Langevin function] can now be said to have
been confirmed by direct experiment.
 Of course, a Langevin function (or any other similar
function) can {\it always} be fitted to such data for a rather narrow
temperature range [citing \cite{giessen,goldfarb}], but the remarkably narrow
 distribution of [\cite{yatsuya}]
 can be
fitted  to such a function over a {\it wide} temperature range. In this
respect this experiment is still quite unique in the literature''
(cf. \cite{barlett}).

Earlier \cite[p. 84]{aharoni}, Aharoni asserts: ``And there is no mistake
in this algebra: there are only two differences between this calculation
[of the Langevin function] and the study of a gas of paramagnetic atoms
in section 2.1 [yielding the Brillouin function]. One is that the function
$\theta$ [the angle to the fixed magnetic field] is continuous here, while
this variable had discrete values in section 2.1 and the other is that the
magnetic moment $\mu$ was that of a single atom there, while here it is the
moment of a large number of atoms, coupled together. However, the second
difference is only quantitative and not qualitative, and the first one
should not make any difference, especially since the energy levels of a large
spin number $S$ are very close together, and look like a continuous variable.
It is thus {\it true} that if there was no other energy term besides the
 isotropic Heisenberg Hamiltonian, it would have been impossible to measure
 any magnetism in zero applied field, and there would be no meaning to
 a Curie termperature, or critical exponents, or any of the other nice
 features mentioned in the
previous chapters. Theorists who calculate these properties never pay
 attention to the fact that the possibility of measuring that which they
 calculate is only due to an extra energy term, which they always leave out. 
Of course, a magnetization as in [the Langevin function], which is zero in
zero applied field, contradicts not only experiments \ldots It is also
in conflict with everyday experience \ldots  It is because
real magnetic materials are not isotropic, and not all values of the angle
$\theta$ are equally probable.''

\subsubsection{Relations to modified Bessel functions}

It is also interesting to observe that the Brillouin and Langevin functions
are both instances ($D=1$ and $D=3$, respectively) of the 
two-spin correlation
function of the $D$-vector model (of isotropically-interacting
$D$-dimensional classical spins) for a one-dimensional lattice
\cite[Fig. 2.3]{stanley},
\begin{equation} \label{stanley}
y_{D} = {I_{{D \over 2}}(J) \over I_{{D \over 2} -1}(J)},
\end{equation}
where $I(x)$ is a modified Bessel function of the first kind.

\subsection{Derivation of the
 Volume Element of the Maximal Monotone Metric for
Spin-1 Systems}
We now discuss the manner in which our first reported result
(\ref{principal}) was derived.
To begin with, we
 noted that Bloore \cite{bloore} had suggested the transformations
(``to suppress the dependence on  [the diagonal entries] $a,b$ and $c$ by
 scaling the [off-diagonal] variables
$f,g,h$''),
\begin{equation} \label{transformations}
f= \sqrt{b c} F, \qquad g= \sqrt{a c} G, \qquad h = \sqrt{a b} H.
\end{equation}
The positivity conditions on the density matrix $\rho$,
 then, took the form \cite[eq. (15)]{bloore},
\begin{equation} \label{positivity}
|F| \leq 1, \qquad |G| \leq 1, \qquad |H| \leq 1, \qquad
|F|^{2} +|G|^2 + |H|^2 -2 \mbox{Re} (FGH) \leq 1.
\end{equation}
Bloore indicated that if $F = F_{R} + \mbox{i} F_{I}$ were simply real
(that is, $F_{I} = 0$), then, the last condition could
be rewritten as
\begin{equation} \label{spheroidal}
1 -F_{R}^{2}  \geq G_{R}^{2} + G_{I}^{2} +H_{R}^{2} +H_{I}^{2}
-2 F_{R} (G_{R} H_{R} -G_{I} H_{I}).
\end{equation}
Then, by the rotation of the
($G_{R},H_{R}$)-pair of axes through an angle of ${\pi \over 4}$ and the
 ($G_{I},H_{I}$)-pair of axes, similarly, the allowed
values of $G_{R}, H_{R},G_{I}, H_{I}$ could
 be seen to lie inside a four-dimensional {\it spheroid}, two of
the principal axes of which had a length equal to $\sqrt{1+F_{R}}$ and
 the other
two, $\sqrt{1-F_{R}}$
(so the spheroid is neither predominantly ``oblate'' nor  ``prolate'' in
character).
Bloore's argument easily extends --- using two additional identical
rotations, not now necessarily however, equal
 to ${\pi \over 4}$  --- to the general case, in which $F$ is
complex. (If we were to single out $G$, say,  rather than $F$, we would
obtain an analogue of (\ref{principal}), containing $b$, not $a$, as its
distinguished  variable among $a,b$ and $c$.)

 We proceed onward by converting
 to polar coordinates, $F_{I} =s \cos{\nu}, F_{R} =s \sin{\nu}$.
Then, we rotate two of the six
 pairs formed by the four axes
 obtained by the two ${\pi \over 4}$-rotations  ---  each
 new pair being comprised of one member from each of the two 
 pairs resulting from the ${\pi \over 4}$-rotations suggested by Bloore.
 The two new angles of rotation now both
equal ${1 \over 2}
\cot^{-1} ({\tan{\nu}})$. 
We have,
 after performing the four indicated rotations,
 transformed the resultant set of axes
($J_{1}, J_{2},J_{3},J_{4}$), using a set of four ``hyperspheroidal''
 coordinates
($r,\theta_{1},\theta_{2},\theta_{3}$), 
\begin{equation} \label{coord}
J_{1} = {r \sqrt{1+s} \cos{\theta_{1}}}, \qquad
J_{2} = {r \sqrt{1+s} \cos{\theta_{2}} \sin{\theta_{1}}},
\end{equation}
\begin{displaymath}
J_{3} = {r \sqrt{1-s} \cos{\theta_{3}} \sin{\theta_{2}} \sin{\theta_{1}}
}, \qquad
J_{4} ={r \sqrt{1-s} \sin{\theta_{3}} \sin{\theta_{2}} \sin{\theta_{1}}}.
\end{displaymath}
The Jacobian of the total transformation (the scaling
transformations (\ref{transformations}), the four rotations of pairs of axes,
along with the introduction of polar and hyperspheroidal coordinates) is,
\begin{equation} \label{jacobian}
J(a,b,c,s,r,\theta_{1},\theta_{2})=
   {a^2 b^2 c^2  r^3 s (1-s^2) \sin{\theta_{1}}^2 \sin{\theta_{2}}}.
\end{equation}
In the new variables, the determinant of the $3 \times 3$
density matrix $\rho$ takes the simple form
(being free of the four angular
 variables --- $u,\theta_{1},\theta_{2},\theta_{3}$),
\begin{equation} \label{determinant}
|\rho| = {a b c (1 -r^2) (1-s^2)}.
\end{equation}

To arrive at the volume element 
 of the maximal monotone metric, which we seek to
integrate over the eight-dimensional convex set of spin-1 density matrices,
we adopted an observation of Dittmann \cite{dittmann}  regarding the
eigenvalues of the sum ($L_{\rho} +R_{\rho}$) of
 the operators of left ($L_{\rho}$) and right multiplication ($R_{\rho}$)
 for the {\it minimal} monotone metric.
He had noted that for the $n \times n$ density matrices, in general, these
$n^2$ eigenvalues would be of the form $p_{i} + p_{j}$ $(i,j =1,\ldots,n)$,
 where the $p_{i}$ $(i=1,\ldots,n)$
 are the
eigenvalues of $\rho$ itself.
Then, we observed \cite{slater2} (cf. \cite[eq. (24)]{hall}) --- making
 use of the fact that
 the determinant of a matrix is
equal to the product of eigenvalues of the matrix --- that
 the corresponding volume element would
be proportional to the {\it square root} of the
 determinant
 of (${L_{\rho} +R_{\rho})^{-1}}$ (or, equivalently,
to the reciprocal of the square root of the
 determinant of $L_{\rho} +R_{\rho}$).
 We adopted this line of argument to the
maximal monotone case by replacing the sums $p_{i}+p_{j}$ by (twice) the 
corresponding {\it harmonic means}
 ($ 2 p_{i} p_{j} /(p_{i}+p_{j})$) \cite{petz}.
Then, the volume element of the maximal monotone metric can be shown
to be proportional to the product of $|\rho|^{-{5 \over 2}}$
 and a  term $w_{2} - |\rho|$, where $w_{2}$ 
denotes \cite{barakat} the sum of the three principal
minors of order two of $\rho$. (The spin-${1 \over
2}$ counterpart of this product --- yielded by the analogous argument, as well
by an independent one \cite[eq. (3.17)]{petzsudar} --- is
simply
${|\rho|^{-{3 \over 2}}}$.)
For the minimal monotone metric in the spin-1 case,
 on the other hand, the volume element --- now based on the terms
 $p_{i} +p_{j}$ --- is
 proportional to $ | \rho |^{-{1 \over 2}}
(w_{2} - | \rho |)^{-1}$.
 The occurrence of
the term $w_{2} - |\rho|$
 in the {\it denominator} of the integrand --- and not the numerator,
as in the spin-1 maximal monotone case --- results in a
substantially more difficult 
multiple integration (cf. (\ref{fullintegration})), for which we have
 been able
to exactly integrate over the variables $u,\theta_{3},\theta_{2}$, leaving
us with an expression containing an elliptic integral of the first kind.
However, further attempts to use numerical methods for this
five-variable expression have so far not been successful.

The spin-${1 \over 2}$ form of the volume element
 for the minimal monotone metric is proportional to
${|\rho|^{-{1 \over 2}}}$. Its use for thermodynamical purposes,
 led \cite{slater5}  to a somewhat different ratio
(cf. \ref{stanley}),
-$I_{2} (\beta) / I_{1}(\beta)$, of modified Bessel functions,
 than the negative of the Langevin
function (\ref{Langevin}), which is expressible \cite{stanley} as
-$I_{3/2}(\beta) / I_{1/2} (\beta)$. (Brody and Hughston \cite{brodyun} had,
 first,
arrived at the former ratio, but then \cite{brody,brody2}
 [personal communication],
 by using
``phase space volume'' rather than ``state density' [i. e., weighted
volume]'', concluded that the latter ratio was more appropriate.
However, for higher dimensional cases --- spin-1, the case under
investigation here, being the simplest
such --- Brody and Hughston \cite{brody} stated that ``the energy surfaces
are not fully ergodic with respect to the Schr\"odinger equation, and
thus we cannot expect to be able to deduce the microcanonical postulate
directly from the basic principles of quantum mechanics.'')
 So, it can be seen that as we pass from the
spin-${1 \over 2}$ case to the spin-1 situation, for both the
minimal and maximal monotone metrics,  we have to deal with an
additional term (besides a certain power of $| \rho |$)
 of the form, $w_{2} -\rho$, to some power.

To find the bivariate marginal probability distribution
 (\ref{principal}), we first performed the
eightfold
multiple integration (using the trace condition to set $c=1-a-b$),
\begin{equation} \label{fullintegration}
\int_0^1 \int_0^{1-a} \int_0^S \int_0^{R} \int_0^{\pi} \int_0^{\pi}
\int_{0}^{2 \pi} \int_{0}^{2 \pi} |\rho|^{-{5 \over 2}} (w_{2} - |\rho|)
J(a,b,c,s,r,\theta_{1},\theta_{2}) \mbox{d} \nu \mbox{d} \theta_{3}
\mbox{d} \theta_{2} \mbox{d} \theta_{1} \mbox{d} r \mbox{d} s 
\mbox{d} b \mbox{d} a.
\end{equation}
Then, we (following established Bayesian principles for
dealing with nonnormalizable prior distributions \cite{bernardo,stone1,stone2})
 took the ratio of the outcome of the (initial sixfold)
 integration --- that is, after all but the
last two stages (over $a$ and $b$) --- to
  the result of the complete integration. In the {\it double} limit,
$R \rightarrow 1, S \rightarrow 1$, the ratio converges to the
probability distribution  (\ref{principal}) over the two-dimensional
simplex. (The result was invariant when the order in which the limits were
taken was reversed.)
 This computational strategy was made necessary due to the fact that the
integration diverged if we directly used the naive
 upper limits, $r = 1$
and $s =1$, evident from (\ref{determinant}). A similar approach  had been
 required  (due to divergence also) in the spin-${1 \over 2}$ case,
based on the maximal monotone metric \cite[eqs. (43)-(46)]{slater1}.
 There, the corresponding volume element ($\propto |\rho|^{-{3 \over 2}}$)
was integrated over a three-dimensional ball of radius $R$. As $R
\rightarrow 1$, this ball coincides with the ``Bloch sphere''
(unit ball) of spin-${1 \over 2}$ systems. A bivariate marginal
probability distribution was then obtainable by taking the limit
$R \rightarrow 1$ of a certain ratio --- with the 
resultant
univariate distribution simply being
uniform in nature, and leading,
 when adopted as
the density-of-states, to the 
 Langevin model (\ref{Langevin}).

Since it has been argued elsewhere \cite{slater3,petztoth} that of all the
possible
monotone metrics, the maximal one is the {\it most noninformative} in
character, it would seem plausible that the maximal metric
 might be singled out to provide density-of-states (structure) functions
for thermodynamic analyses. (For additional results pertinent  --- in
the context of ``universal quantum coding'' --- to comparative
properties of the maximal and monotone metrics, see \cite{kratten}.)

We would also like to report a success
 in performing the eightfold multiple integration
in (\ref{fullintegration}), after reordering the individual integrations so
that the ones over $r$ and $s$ --- rather than $u$ and $\theta_{3}$ --- were
 performed first. Then, we were able to
compute the limit of the ratio of the result after the first {\it two}
 integrations
to the complete eightfold
 integration. This gives us a {\it six}-dimensional marginal
 probability distribution (free of $r$ and $s$, and
 independent of the particular values of $\nu$ and
$\theta_{3}$),
\begin{equation} \label{end}
{15  (1-a) \sqrt{a} \sin{\theta_{1}}^{4} \sin{\theta_{2}}^{3} \over 8 
{\pi}^{4} \sqrt{b} \sqrt{c}}.
\end{equation}
One of the two-dimensional marginal probability distributions
 of (\ref{end}) --- the one over the simplex spanned by the three
diagonal entries ($a,b,c$) of the density matrix (\ref{blooremat}) --- is
 our
earlier result (\ref{principal}).

The focus of the study here has been on the application
 of {\it symbolic} integration to the modeling of the thermodynamics
properties of spin-1 systems. With the application of {\it numerical}
integration methods, however, we might hope to broaden our work to encompass
{\it non}-diagonal Hamiltonians (such as 
 $\lambda_{1}, \lambda_{2},\lambda_{4},
\lambda_{5},\lambda_{6},\lambda_{7}$).
\subsection{Thermodynamic Analysis Based on a {\it Non-Diagonal} Hamiltonian
 ($\lambda_{1}$)} \label{nondiagonal1}
\begin{figure}
\centerline{\psfig{figure=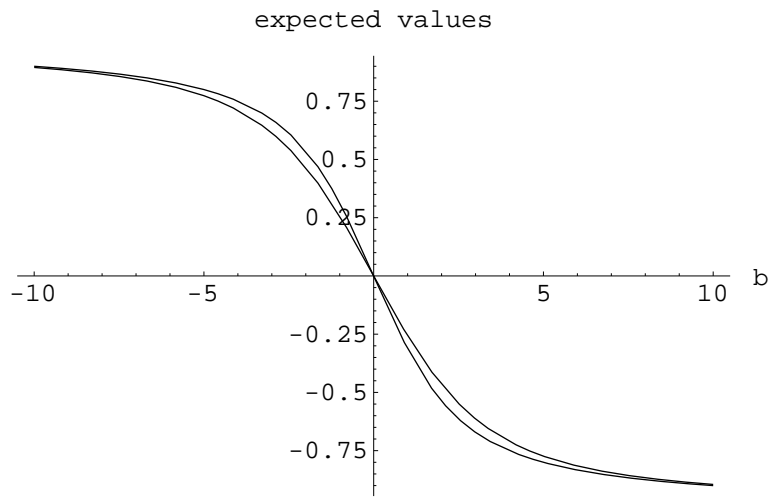}}
\caption{Expected value of $\langle \lambda_{1} \rangle$ and the negative
of the Langevin function (\ref{Langevin})}
\label{almostlangevin}
\end{figure}

\begin{figure}
\centerline{\psfig{figure=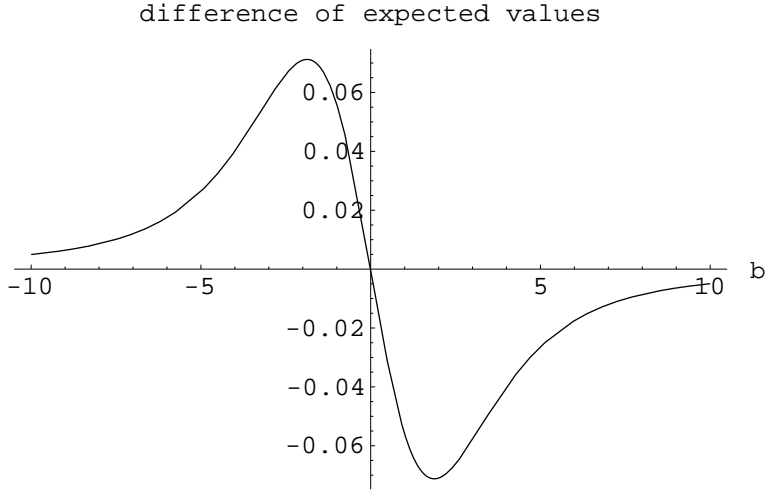}}
\caption{Difference between the negative of the Langevin function
(\ref{Langevin}) and $\langle \lambda_{1} \rangle$}
\label{diffcurve}
\end{figure}

Let us proceed with a ``strong equilibrium'' analysis \cite{park}
of the non-diagonal observable,
\begin{equation} \label{nondiagonal}
 \lambda_{1} =  \pmatrix{0 & 1 & 0 \cr
                       1 & 0 & 0 \cr
                       0 & 0 & 1 \cr}.
\end{equation}
(It would appear that all our conclusions of a thermodynamic nature should
be precisely the same for the other five non-diagonal observables, as well.)
Rather than considering the full eight-dimensional convex set of
$3 \times 3$ density matrices, however, we restrict our considerations
to the two-dimensional convex set of $3 \times 3$ density matrices which
{\it commute} with $\lambda_{1}$ --- so, we are considering
the case of ``strong equilibrium'' \cite{park}.
 We, then, assign the volume element of
the Bures (minimal monotone)
 metric to this set (using formula (10) in \cite{hub1}), as a
density-of-states.
 (Since the set is composed of commuting matrices,
we are in a {\it classical} situation in which there is a {\it unique}
 monotone metric \cite{petzsudar},
so there is, in fact,
 no distinction between maximal and minimal ones, as earlier.)
Our partition function then takes the form,
\begin{equation}
Q(\beta) = \int_{0}^{1} \int_{0}^{1-g} {e^{-\beta
 (1 -2 g - h)} \over {2 \sqrt{g} \sqrt{h} \sqrt{1-g-h}}} \mbox{d} h \mbox{d} g.
\end{equation}
(We perform the inner integration symbolically, and the outer one,
numerically.)
In Fig.~\ref{almostlangevin} (cf. Fig.~\ref{LLang}),
 we plot the associated expected value, along
with that given by the negative of the Langevin function (\ref{Langevin}).
Since these two curves quite remarkably almost coincide, in
 Fig.~\ref{diffcurve} we plot the result obtained by subtracting from the
negative of the Langevin function (\ref{Langevin}), the expected value of
$\lambda_{1}$.
\section{SPIN-${3 \over 2}$ OR COUPLED SPIN-${1 \over 2}$ 
SYSTEMS} \label{spinspin}
\subsection{Required Transformations of the Fifteen Parameters}
The possibility of extending our line of analysis
to the $(n^{2} -1)$-dimensional convex sets of $n \times n$ density matrices,
for $n > 3$, obviously,
remains to be fully investigated.
The $3 \times 3$ density matrix (\ref{blooremat}) can be considered to be
embedded in the upper left corner of the $n \times n$ density matrix.
Then, for this block, we can employ precisely the same set of transformations
as in the analysis above. (Of course, there are ${n (n-1) \over 3}$ ways in
which to actually
 position the $3 \times 3$ block. Depending upon which we choose,
we would expect to break certain symmetries between the diagonal entries,
as we witnessed with the probability distribution (\ref{principal}).)
This leaves us with $n-3$ new diagonal entries and
$n^{2}-10$ new off-diagonal variables.

 For the case $n=4$
(which we had previously studied \cite{slater1} for various scenarios
in which most of the fifteen parameters were {\it ab initio} set to zero),
\begin{equation} \label{bleeremat}
\rho = \pmatrix{a & \bar{h} & g & \bar{o} \cr
 h & b & \bar{f} & p \cr
 \bar{g} & f & c & \bar{q} \cr
o & \bar{p} & q & d \cr}, \qquad a, b, c, d \in {\Bbb{R}},
\qquad f, g, h, o, p, q \in{\Bbb{C}}
\end{equation}
 we have found that in the determinant,
the cross-product between the real and complex parts of each new (scaled)
 off-diagonal entry (that is, $o,p,q$) is zero.
 Then, the extension of our earlier
reasoning led us to form two {\it triads}, one element of each triad coming
from one of these
three pairs, and attempt to
 rotate them (in three-dimensional space), so as to
eliminate the non-zero cross-product terms \cite{smith,boj,hoff,mallesh}.
We have also 
investigated the
 possibility that  by rotating the six variables in question in
six-dimensional space, one could, in a single
process, nullify all the non-zero cross-product terms.
To accomplish this, it would be necessary to diagonalize a $6 \times 6$ 
symmetric matrix (or, equivalently, a $3 \times 3$ Hermitian matrix).
The three pairs of
equal
eigenvalues of this matrix are given by
 the solution of the  {\it cubic} equation,
\begin{equation} \label{cubiceq}
x^3 + {1 \over 2} ((6 -2 r^2 -r^2 s -2 s^2 -r^2 s
(1 - 4 \sin{\theta_{1}}^2 \sin{\theta_{2}}^2)) x^2 
+3 (1-r^2) (1-s^2) x + (1-r^2)^2  (1-s^2)^2  = 0.
\end{equation}
There does not appear to be any very 
 simple general form for the three roots of this
equation. (In principle, we are able to diagonalize the
$6 \times 6$ matrix, but the computations required to fully implement the
associated transformations of the six variables, seem to be quite daunting.)
However, if we set the factor ($1- 4 \sin{\theta_{1}}^2
\sin{\theta_{2}}^2$) to zero (which can be accomplished by taking
 either $\theta_{1}$ or $\theta_{2}$ to be zero), then the
three roots are of the form
\begin{equation} \label{cubicroots}
x =(-1+r^2) (1+s),\qquad
x = {1 \over 2} (1-s) (-2 -s + p), \qquad x = -{1 \over 2} (1-s) (2 +s +p),
\end{equation}
where 
\begin{equation}
p = \sqrt{s^2 +4 r^2 (1+s)}.
\end{equation}
If we set the factor to be -3 (by taking both $\theta_{1}$ and $\theta_{2}$
to be ${\pi \over 2}$), then the roots are
\begin{equation} \label{cubicroots2}
x= (-1 + r^2) (1- s),\qquad x = {1 \over 2} (1+s) (-2+s +p),\qquad
x=-{1 \over 2} (1+s) (2-s+p),
\end{equation}
where now
\begin{equation}
p=\sqrt{s^2 +4 r^2 (1-s)}.
\end{equation}
The situation becomes simpler still if we set the parameter $s$ to zero
(so that the $2 \times 2$ submatrix of (\ref{blooremat}) in the upper left
corner is diagonal in nature).
Then, the three roots are  $-1-r$,$-1+r$ and $-1+r^2$.
Similarly, for $r=0$, the roots are $-1-s$,$-1+s$ and $-1+s^2$.
If we set $s=1$ (corresponding to a pure state), two of the roots are zero
and the third is $-2 (1- r^2 +r^2 \sin{\theta_{1}}^2 \sin{\theta_{2}}^2$).
For $r=1$, there are also two zero roots and one equalling 
$-2 +s +s^2 -2 s \sin{\theta_{1}}^2 \sin{\theta_{2}}^2$.

Let us proceed to analyze in detail
 the case $r=0$, for which the three eigenvalues, as
just noted, are $-1-s$,$-1+s$ and $-1+s^2$. Setting $r$ to zero is
equivalent (cf. (\ref{coord}))
 to nullifying the entries $g$ and $h$ in (\ref{bleeremat}),
so the scenario is  eleven-dimensional in nature.
We have found the eigenvectors corresponding to these three eigenvalues,
that is, for ($-1-s$),
\begin{equation} \label{eigenvectors}
(0,0,{\cos{\nu} \over \sqrt{2}},-{\sin{\nu} \over
 \sqrt{2}},0,{1 \over \sqrt{2}}),\qquad  (0,0,{\sin{\nu} \over \sqrt{2}},
{\cos{\nu} \over \sqrt{2}},{1 \over \sqrt{2}},0),
\end{equation}
for ($-1+s$),
\begin{displaymath}
(0,0,-{\cos{\nu} \over \sqrt{2}},{\sin{\nu} \over
 \sqrt{2}},0,{1 \over \sqrt{2}}),\qquad (0,0,-{\sin{\nu} \over \sqrt{2}},
-{\cos{\nu} \over \sqrt{2}},{1 \over
\sqrt{2}},0),
\end{displaymath}
and for ($-1+s^2$),
\begin{displaymath}
(0,1,0,0,0,0), \qquad (1,0,0,0,0,0).
\end{displaymath}
We transformed the three complex variables
 $o,p$ and $q$, on the basis of these eigenvectors,
 to a new set of six real variables ($K_{1}\ldots,K_{6}$),
between which there are no nonzero cross-product terms in the determinant
of the transformed matrix.
Then, we employed a six-dimensional version of spheroidal coordinates
($v,\xi_{1},\xi_{2},\xi_{3},\xi_{4},\xi_{5}$) (cf. (\ref{coord})),
\begin{equation} \label{spheroidal6}
K_{1}= v \sqrt{1+s} \cos{\xi_{1}}, \qquad K_{2} = v \sqrt{1+s} \cos{\xi_{2}}
 \sin{\xi_{1}},
\end{equation}
\begin{displaymath}
 K_{3} = v \sqrt{1-s} \cos{\xi_{3}} \sin{\xi_{2}} \sin{\xi_{1}},\qquad
 K_{4} =v \sqrt{1-s} \cos{\xi_{4}} \sin{\xi_{3}} \sin{\xi_{2}} \sin{\xi_{1}},
\end{displaymath}
\begin{displaymath}
 K_{5} =v \cos{\xi_{5}} \sin{\xi_{4}} \sin{\xi_{3}}
 \sin{\xi_{2}} \sin{\xi_{1}},\qquad
 K_{6}= v \sin{\xi_{5}} \sin{\xi_{4}} \sin{\xi_{3}}
 \sin{\xi_{2}} \sin{\xi_{1}}.
\end{displaymath}
 The determinant becomes, then,
simply $a b c d (1-s^2) (1-v^2)$. The Jacobian corresponding to the
sequence of  transformations is (cf. (\ref{jacobian}))
\begin{equation} \label{jacobian2}
J(a,b,c,d,s,v,\xi_{1}, \xi_{2}, \xi_{3},\xi_{4})
 =a b^2 c^2 d^3 s (1-s^2) v^5  \sin{\xi_{1}}^4 \sin{\xi_{2}}^3
\sin{\xi_{3}}^2 \sin{\xi_{4}}.
\end{equation}
\subsection{Volume Elements of the Maximal and Minimal Monotone Metrics
for Spin-${3 \over 2}$ Systems}
The integrand based on the maximal monotone metric is (employing 
once again our {\it ansatz} based on the work of Dittmann \cite{dittmann},
in which we replace the {\it arithmetic} means
of the eigenvalues by their {\it harmonic} means, in accordance with
the corresponding Morozova-Chentsov functions),
 the product
of this Jacobian and the ratio of the term
($w_{2} w_{3} -w_{3}^2 -|\rho|$) to $|\rho|^{{7 \over 2}}$, where
$w_{2}$ is the sum of the six principal minors of order two and
$w_{3}$ is the sum of the four principal minors of order three.
We performed the integration over the eleven-dimensional convex set, the
innermost integrations being over $v$ from 0 to $V$, and $s$ from
0 to $S$. (The integration would
diverge if we were simply to set either $S$ or $V$ to 1.)
Then, we were able to integrate over the six angular variables ($u,\xi$'s),
but not fully over the three-dimensional simplex spanned by the diagonal
entries ($a,b,c,d$). 

The problem, in this regard, seems attributable to the fact that the powers
of $a,b$ and $c$ --- but not $d$ --- in the Jacobian (\ref{jacobian2})
 are all less than three. If all four powers were, in fact,
 equal to three, then dividing by $|\rho|^{7 \over 2}$,
with $|\rho| \propto a b c d$,
would yield a term of the form $(a b c d)^{-{1 \over 2}}$ --- similar to
the term $(a b c)^{-{1 \over2}}$, encountered in the successful
spin-1 analysis above --- which would be
integrable over the simplex. (We would, in fact, obtain such a desirable
term if we could otherwise proceed with a {\it full}
fifteen-dimensional analysis.) However, we have the resultant term
$ a^{-{5 \over 2}} b^{-{3 \over 2}} c^{-{3 \over 2}} d^{-{1 \over 2}}$,
which appears to be not integrable.
Thus, we have not so far been able to extend the form of analysis
taken here for the spin-1 systems to the convex set of
spin-${3 \over 2}$ or, equivalently, coupled spin-${1 \over 2}$ systems
(cf. \cite{slater1}). (Following the lead of Bloore \cite{bloore}, one might
also study the presumably simpler case of the nine-dimensional 
convex set of {\it real} $4 \times 4$ density
matrices.)

\subsection{{\it Strong} Equilibrium Thermodynamic
 Analysis of a {\it Non-Diagonal} Hamiltonian}

\begin{figure}
\centerline{\psfig{figure=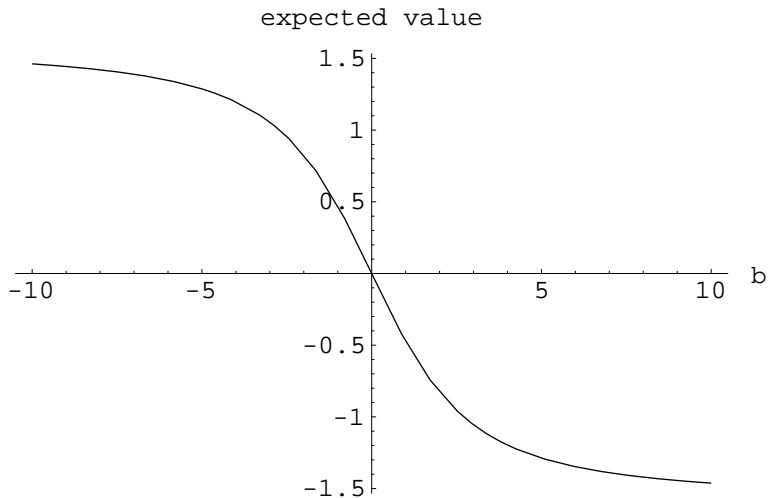}}
\caption{Expected value of the non-diagonal observable (\ref{observable4by4})}
\label{plot4by4}
\end{figure}

Let us, in the manner of sec.~\ref{nondiagonal1},
 restrict our attention to those
$4 \times 4$ density matrices which commute with a certain non-diagonal
Hamiltonian (so we remain within the framework of
``strong equilibrium''), which we choose here to be
\begin{equation} \label{observable4by4}
\pmatrix{0 & 1 & 0 & 0\cr
         1 & 0 & 1 & 0\cr
         0 & 1 & 0 & 1\cr
         0 & 0 & 1 & 0\cr}.
\end{equation}
Then, after assigning the volume element of the Bures metric to the
three-dimensional convex set of the mutually commuting density matrices,
we have found the expected value of the observable (\ref{observable4by4}),
as a function of the thermodynamic parameter $\beta$,
to be as indicated in Fig.~\ref{plot4by4}.

\section{CONCLUDING REMARKS}

In this communication, we have sought to pursue a particular chain
of reasoning (which, for spin-${1 \over 2}$ systems, is known to
 yield the Langevin function, thus according fully with detailed
arguments of Lavenda \cite{lavenda})
 for spin-1 and spin-${3 \over 2}$ systems.
The fundamental problems encountered 
{\it en route} include the derivation of the volume
elements of the (noninformative) maximal monotone metric 
(of the left logarithmic derivative) for these 
two scenarios. By applying principles used in Bayesian reasoning
\cite{bernardo,stone1,stone2}, we have
succeeded through extensive analysis and computations in obtaining a 
six-dimensional marginal probability distribution (\ref{end}) from the
(nonnormalizable) volume
element for the eight-dimensional convex set of spin-1 systems
(sec.~\ref{spin1systems}).
We have not been quite as
 successful, however, in addressing the spin-${3 \over 2}$
scenario, in which the corresponding convex set is fifteen-dimensional
in nature (sec.~\ref{spinspin}).

We are not able to actually demonstrate, but can only conjecture, that
 our results have physical implications (cf. \cite{brody}).
The most questionable and
debatable assumption underlying our analysis, perhaps, is that
the volume element of the maximal monotone metric can be employed as
a density-of-states for thermodynamic purposes.

The initial impetus to pursue this line of research 
was provided by
an extended series of forcefully-argued
 papers of Band and Park \cite{park}. In them, for various 
conceptual reasons, they
expressed dissatisfaction with the conventional (``Jaynesian'')
approach to quantum statistical thermodynamics and sought to develop
a conceptually superior  alternative approach.
 We believe that our
analysis is quite consistent with
 the spirit of their work, and serves to implement
it for  certain low-dimensional scenarios.
We have also been encouraged in our  work by the detailed remarks of
Lavenda \cite{lavenda}, regarding the propriety of using the Langevin
function for spin-${1 \over 2}$ systems, since our methodology, in fact,
yields this function.
Also, the analyses of Brody and Hughston \cite{brody,brody2},
 though differing from
ours in several technical respects --- in particular, focusing on
pure states --- possess similar
motivations.

It also appears natural that, amongst the {\it continuum} of monotone metrics
\cite{petz0,petzsudar}, the {\it maximal} one should play the distinguished
information-theoretic role we have explored for it,
 since its {\it noninformative}
nature has been independently
established  \cite{slater3,petztoth}.
The fact that, as a result of lengthy and demanding computations
based on the volume element of the maximal monotone metric,
we were able to arrive at the quite simple results (\ref{principal}) 
and (\ref{end}) also assists, we believe, in validating the rather
unconventional course pursued here.

\acknowledgments

I would like to express appreciation to the Institute for Theoretical
Physics for computational support in this research and to Christian
Krattenthaler for an illuminating discussion.

\listoffigures

\end{document}